\documentclass{sig-alternate}

\newcommand{\x}{\mathbf{x}}
\newcommand{\w}{\mathbf{w}}
\newcommand{\z}{\mathbf{z}}

\newcommand{\q}{\mathbf{q}}

\newcommand{\Wb}{\mathbf{W}}

\newcommand{\Qb}{\mathbf{Q}}

\newcommand{\Xb}{\mathbf{X}}
\newcommand{\Sb}{\mathbf{S}}
\newcommand{\Rb}{\mathbf{R}}

\newcommand{\sname}{dCoT} 
\newcommand{\snamelong}{Dense Cohort of Terms}
\usepackage{url}

\begin{document}
%
\conferenceinfo{CIKM'12,} {October 29--November 2, 2012, Maui, HI, USA.} 
\CopyrightYear{2012} 
\crdata{978-1-4503-1156-4/12/10} 
\clubpenalty=10000 
\widowpenalty = 10000

\title{An alternative text representation \\
to TF-IDF and Bag-of-Words
\titlenote{A modified version of our CIKM 2012 paper: \emph{From sBoW to dCoT, Marginalized Encoders for Text Representation}}}

%
%
%
%
%

\numberofauthors{4} 
%
\author{
%
%
\alignauthor
Zhixiang (Eddie) Xu
       \affaddr{Washington University in St. Louis}\\
       \affaddr{One Brookings Dr.}\\
       \affaddr{St. Louis, MO, USA}\\
       \email{zhixiang.xu@wustl.edu}
\alignauthor
Minmin Chen \\
       \affaddr{Washington University in St. Louis}\\
       \affaddr{One Brookings Dr.}\\
       \affaddr{St. Louis, MO, USA}\\
       \email{chenm@wustl.edu}
\alignauthor 
Kilian Q. Weinberger
       \affaddr{Washington University in St. Louis}\\
       \affaddr{One Brookings Dr.}\\
       \affaddr{St. Louis, MO, USA}\\
       \email{kilian@wustl.edu}
\and  
\alignauthor 
Fei Sha\\
       \affaddr{University of Southern California}\\
       \affaddr{941 West 37th Place}\\
       \affaddr{Los Angeles, CA, USA}\\
       \email{feisha@usc.edu}
}

\date{31 July 2012}
\sloppy
\maketitle
\begin{abstract}

In text mining, information retrieval, and machine learning, text documents are commonly represented through variants of \emph{sparse Bag of Words} (sBoW) vectors (\emph{e.g.} TF-IDF~\cite{jones1972statistical}). Although simple and intuitive, sBoW style representations suffer from their inherent over-sparsity and fail to capture word-level synonymy and polysemy. 
Especially when labeled data is limited (\emph{e.g.} in document classification), or the text documents are short (\emph{e.g.} emails or abstracts), many features are rarely observed within the training corpus. This leads to overfitting and reduced generalization accuracy. In this paper we propose \emph{\snamelong{}} (\sname{}), an unsupervised algorithm to learn improved sBoW document features. \sname{} explicitly models absent words by removing and reconstructing random sub-sets of words in the unlabeled corpus. 
With this approach, \sname{} learns to reconstruct frequent words from co-occurring infrequent words and maps the high dimensional sparse sBoW vectors into a low-dimensional dense representation. We show that the feature removal can be marginalized out and that the reconstruction can be solved for in closed-form. 
We demonstrate empirically, on several benchmark datasets, that \sname{} features significantly improve the classification accuracy across several document classification tasks.


\end{abstract}




\section{Introduction}
\label{sec:intro}
The feature representation of text documents plays a critical role in many applications of data-mining and information retrieval. The \emph{sparse Bag of Words} (sBoW) representation is arguably one of the most commonly used and effective approaches. Each document is represented by a high dimensional sparse vector, where each dimension corresponds to the term frequency of a unique word within a dictionary or hash-table~\cite{weinberger2009feature}. A natural extension is TF-IDF~\cite{jones1972statistical}, where the term frequency counts are discounted by the inverse-document-frequencies.  Despite its wide-spread use with text and image data~\cite{csurka2004visual}, sBoW does have some severe limitations, mainly due to its often excessive sparsity. 

Although the Oxford English Dictionary contains approximately $600,000$ unique words, it is fair to say that the essence of most written text can be expressed with a far smaller vocabulary (\emph{e.g.} $5000\!-\!10000$ unique words). For example the words \emph{splendid, spectacular, terrific, glorious, resplendent} are all to some degree synonymous with the word \emph{good}. However, as sBoW does not capture synonymy, 
a document that uses ``splendid'' will be considered dissimilar from a document that uses the word ``terrific''. 
A classifier, trained to predict the sentiment of a document, would have to be exposed to a very large set of labeled examples to learn that all these words are predictive towards a positive sentiment.

In this paper, we propose a novel feature learning algorithm that directly addresses the problems of excessive sparsity in sBoW representations. Our algorithm, which we refer to as \snamelong{} (\sname{}), maps high-dimensional (overly) \emph{sparse} vectors into a low-dimensional \emph{dense} representation.
The mapping is trained to reconstruct frequent from \emph{in}frequent words. The training process is entirely unsupervised, as we generate training instances by randomly and repeatedly removing common words from text documents. These removed words are then reconstructed from the remaining text. In this paper we show that the feature removal process can be marginalized out and the reconstruction can be solved for in closed form. The resulting algorithm is a closed-form transformation of the original sBoW features, which is extremely fast to train (on the order of seconds) and apply (milliseconds).

Our empirical results indicate that \sname{} is useful for several reasons. First, it provides researchers with an efficient and convenient method to learn better feature representation for sBoW documents, and can be used in a large variety of data-mining, learning and retrieval tasks.
Second, we demonstrate that it clearly outperforms existing document representations~\cite{jones1972statistical,blei2003latent,deerwester1990indexing} on several classification tasks. Finally, it is much faster than most competing algorithms. 


\section{Related Work}
\label{sec:related}

Over the years, a great number of models have been developed to describe textual corpora, including vector space models~\cite{luhn1957statistical, salton1975vector, salton1988term, singhal1996pivoted,Weinberger08large}, and topic models~\cite{blei2003latent,hofmann1999probabilistic, Nigam99textclassification, bai09supervised}. Vector space models reduce each document in the corpus to a vector of real numbers, each of which reflects the counts of an unordered collection of words. Among them, the most popular  one is the TF-IDF scheme~\cite{salton1988term}, where each dimension of the feature vector computes the term frequency count factored by the inverse document frequency count. By down-weighting terms that are common in the entire corpus, it effectively identifies a subset of terms that are discriminative for documents in the corpus. Though simple and efficient, TF-IDF reveals little of the correlations between terms, thus fails to capture some basic linguistic notions such as synonymy and polysemy. Latent Semantic Index (LSI)~\cite{deerwester1990indexing} attempts to overcome this. It applies Singular Value Decomposition (SVD)~\cite{golub1970singular} to the TF-IDF (or sBoW) features to find a so-called latent semantic space  that retains most of the variances in the corpus. Each feature in the new space is a linear combination of the original TF-IDF features, which naturally handles the synonymy problem. 

Topic modeling develops generative statistical models to discover the hidden ``topic'' that occur in the corpus. Probabilistic LSI~\cite{hofmann1999probabilistic}, which is proposed as an alternative to LSI, models each document as a mixture of a fixed set of topics, and each word as a sample generated from a single topic. The limitation of probabilistic LSI is that the mixture of topics is modeled explicitly for each training data using a large set of individual parameters, hence, there is no natural way to assign probabilities to unseen documents.   Latent Dirichlet Allocation (LDA)~\cite{blei2003latent}  solves the problem by introducing a Dirichlet prior on the topic distribution, and treating the mixing weights as multinomial distributed random variables. It is probably the most commonly used topic models nowadays, and the posterior Dirichlet parameters are often used as the low dimensional representation for various tasks~\cite{blei2003latent}. \cite{blitzer05hierarchical} use non-linear dimensionality reduction~\cite{weinberger04learning} to embed text data into a low dimensional space, while preserving pair-wise distances between documents. It is fair to say that their approach is computationally most demanding. Similarly to LSI, pLSI and LDA, our algorithm also maps the sparse sBoW features into a low dimensional dense representation. However it is faster to train and addresses the problem of synonymy more explicitly. 


\section{\sname{}}
\label{sec:method}

First, we introduce notations that will be used throughout the paper. Let $D = \{\w_1, \cdots, \w_d\}$ be the dictionary of words that appear in the text corpus, with size $d=|D|$. Each input document is represented as a vector $\x \in {\cal R}^d$, where each dimension $x_j$ counts the appearance of word $\w_j$ in this document. Let $\mathbf{X} = [\x_1, \cdots, \x_n]$ denote the corpus. Assume that the first $n_l \ll n$ documents are accompanied by corresponding labels $\{ y_1, \cdots, y_{n_l}\} \in \cal Y$, drawn from some joint distribution $\cal D$. 

In this section we introduce the algorithm \sname{}, which translates sparse sBoW vectors $\x\in{\cal R}^d$ into denser and lower dimensional prototype vectors. We first define the concept of \emph{prototype} terms and then derive the algorithm step-by-step. 

\textbf{Prototype features.}
Let $P = \{\w_{p_1}, \cdots, \w_{p_r}\} \subset D$, with $|P|=r$ and $r \ll d$, denote a strict subset of the vocabulary $D$, which we refer to as \emph{prototype} terms. 
Our algorithm aims to ``translate'' each term in $D$ into one or more of these prototype words with similar meaning. 
Several choices are possible to identify $P$, 
but a typical heuristic is to pick the $r$ most frequent terms in $D$. The most frequent terms can be thought of as representative expressions for sets of synonyms --- \emph{e.g.} the frequent word \emph{good} represents the rare words \emph{splendid, spectacular, terrific, glorious}. For this choice of $P$, \sname{} translates \emph{rare} words into \emph{frequent} words. 


\textbf{Corruption.} 
The goal of \sname{} is to learn a mapping $\Wb:{\cal R}^d\rightarrow{\cal R}^r$, which ``translates'' the original sBoW vectors in ${\cal R}^d$ into a combination of prototype terms in ${\cal R}^r$. 
Our training of $\Wb$ is based on one crucial insight: If a prototype term already exists in some input $\x$, $\Wb$ should be able to predict it from the remaining terms in $\x$. We therefore artificially create a supervised dataset from unlabeled data by \emph{removing} (\emph{i.e.} setting to zero) each term in $\x$ with some probability $(1-p)$. We perform this $m$ times and refer to the resulting corrupted vectors as $\hat{\x}^1,\dots,\hat{\x}^m$. We not only remove prototype features but all features, to generate more diverse input samples. (In the subsequent section we will show that in fact we never actually have to create this corrupted dataset, as its creation can be marginalized out entirely --- but for now let us pretend it actually exists.)

\textbf{Reconstruction.} In addition to the corruptions, for each input  $\x_i$ we create a sub-vector $\bar\x_i =[x_{p_1}, \cdots, x_{p_r}]^\top \in {\cal R}^r$  which only contains its prototype features. A mapping $\Wb \in {\cal R}^{r \times d}$ is then learned to reconstructs the prototype features from the corrupted version $\hat{x}_i$, by minimizing the squared reconstruction error,
\begin{equation}
	\frac{1}{2nm}\sum_{i=1}^n\sum_{j=1}^m \|\bar\x_i - \Wb\hat{\x}_i^j\|^2. \label{eq:msl}
\end{equation}

To simplify notation, we assume that a constant feature is added to the corrupted input, $\hat{\x}_i = [\hat{\x}_i ; 1]$, and an appropriate bias is incorporated within the mapping $\Wb = [\Wb,\mathbf{b}]$. Note that the constant feature is never corrupted. The bias term has the important task of reconstructing the average occurrence of the prototype features.  


Let us define a design matrix 
\begin{equation}
\overline{\Xb} = [\underbrace{\bar\x_1, \cdots, \bar\x_1}_{m}, \cdots, \underbrace{\bar\x_n, \cdots, \bar\x_n}_{m}] \in {\cal R}^{r\times nm}	\nonumber
\end{equation}
 as the $m$ copies of the prototype features of the inputs. Similarly, we denote the $m$ corruptions of the original inputs as $\widehat{\Xb} = [\underbrace{\hat{\x}_1^1, \cdots, \hat{\x}_1^m}_{m}, \cdots, \underbrace{\hat{\x}_n^1, \cdots, \hat{\x}_n^m}_{m}] \in {\cal R}^{d\times nm}$. With this notation, the loss in eq.~(\ref{eq:msl}) reduces to
\begin{equation}
	\frac{1}{2nm} \|\overline{\Xb} - \Wb \widehat{\Xb}\|_F^2, \label{eq:matrix}
\end{equation}
where $\|\cdot\|_F^2$ denotes the squared Frobenius norm. 
The solution to~(\ref{eq:matrix}) can be obtained under closed-form as the solution to the well-known ordinary least square.
\begin{equation}
	\Wb = \Rb \Qb^{-1} \textrm{ with } \Qb=\widehat \Xb\widehat \Xb^{\top} \textrm{ and } \Rb=\overline{\Xb} \widehat \Xb ^{\top}.	\label{eq:solution}
\end{equation}

\textbf{Marginalized corruption. }
Ideally, we would like the number of corrupted versions become very large, \emph{i.e.} $m \rightarrow \infty$. By the weak law of large numbers, $\Rb$ and $\Qb$ then converge to their expectations and (\ref{eq:solution}) becomes
\begin{equation}
	\Wb=E[\Rb]E[\Qb]^{-1},\label{eq:Wlimit}
\end{equation}
with the expectations of $\Rb$ and $\Qb$ defined as
\begin{equation}
	E[\Qb] = \sum_{i=1}^nE[\hat{\x}_i\hat{\x}_i^{\top}],\textrm{ }	E[\Rb] = \sum_{i=1}^nE[\bar{\x}_i\hat{\x}_i^{\top}].\label{eq:expectations}
\end{equation}
The uniform corruption allows us to compute the expectations in (\ref{eq:expectations}) in closed form. 
%
%
Let us define a vector $q=[p,\dots,p,1]^\top \in {\cal R}^{d+1}$, where $q_{\alpha}$ indicates if feature $\alpha$ survive a  corruption (the constant feature is never corrupted, hence $q_{d+1}\!=\!1$). If we denote the scatter matrix of the uncorrupted input as $S = \Xb \Xb^{\top}$, we obtain $E[\Rb]_{\alpha\beta}	 = \Sb_{\alpha\beta} \q_\alpha$ and 
\begin{equation}
E[\Qb]_{\alpha\beta}	 = \left\{  
	\begin{array}{lcc}
		\Sb_{\alpha\beta} \q_\alpha \q_\beta & \mbox{if} & \alpha\neq \beta \\
		\Sb_{\alpha\beta} \q_\alpha   & \mbox{if} & \alpha=\beta \label{eq:expected_w}
	\end{array}.\right.
\end{equation}
The diagonal entries of $E[\Qb]$ are the product of two identical features, and the probability of a feature surviving  corruption is $p$. The expected value of the diagonal entries is therefore the scatter matrix multiplied by $p$. 
The off-diagonal entries are the product of two different features $\alpha$ and $\beta$, which are corrupted independently. The probability of both features surviving the corruption is $p^2$.

\textbf{Squashing function.} The output of the linear mapping $\Wb:{\cal R}^d\rightarrow {\cal R}^r$ approximates the expected value~\cite{Bishop06} of a prototype term. It can be beneficial to have more bag-of-word like features that are either present or not. For this purpose, we apply the $\tanh()$ squashing-function to the output 
\begin{equation}
	\z=\tanh(\Wb\x), \label{eq:madoff}
\end{equation}
which has the effect of amplifying or dampening the feature values of the reconstructed prototype words. 
We refer to our feature learning algorithm as \sname{} (\snamelong{}).

\begin{figure}[t!!!]
	\vspace{-1.5ex}
\centerline{
\includegraphics[width = .9618\columnwidth]{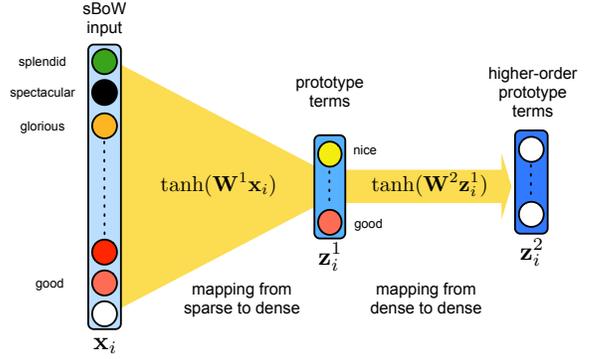}
}
\vspace{-1.5ex}
\caption{Schematic layout of  \sname{}. The left part illustrates that \sname{} learns a mapping from the overly sparse BoW representation to a dense one. 
The right part illustrates the recursive re-application to reconstruct prototype features from the context.}
\label{fig:sparse}
\end{figure}




\subsection{Recursive re-application}

The linear mapping in eq.(\ref{eq:madoff}) is trained by reconstructing prototype words from partially corrupted input vectors. This linear approach works well for prototype words that commonly appear together with words of similar meaning (\emph{e.g.} ``president'' and ``obama''), as the mapping captures the correlation between the two. It can however be the case that two synonyms never appear together because the input documents are short and the authors use one term or the other but rarely both together (\emph{e.g.} ``tasty'' and its rarer synonym ``delicious''). 
%
%
%
In these cases it can help to recursively re-apply \sname{} to its own output. Here, the first mapping reconstructs a common context between synonyms (\emph{i.e.} words that co-occur with all synonyms) and subsequent applications of \sname{} reconstruct the synonym-prototypes from this context. In the previous example, one could imagine that the first application of \sname{} constructs context prototype words like ``food'', ``expensive'', ``dinner'', ``wonderful'' from the original term ``delicious''. The re-application of \sname{} reconstructs ``tasty'' from these context words. 

Let the mapping from eq.(\ref{eq:madoff}) be $\Wb^1\in{\cal R}^{r\times d}$ and $\z^1_i=\tanh\left(\Wb^1\x_i\right)$, for an input $\x_i$. We now compute a second mapping $\Wb^2\in{\cal R}^{r\times r}$, exactly as defined in the previous section, except that  we consider the vectors $\z^1_1,\dots,\z^1_n\in{\cal R}^r$ as input. The mapping $\Wb^2$ is an affine transformation which stays within the prototype space spanned by $P$. This process can be repeated many times and because the input dimensionality is low the computation of (\ref{eq:madoff}) is cheap. Figure~\ref{fig:sparse} illustrates this process in a schematic layout. 

If \sname{} is applied $l$ times, the final representation $\z_i$ is the concatenated vector of all outputs and  the original input, 
\begin{equation}
\z_i = ( \x_i,\z_i^1,\cdots,\z_i^l)^\top.
\end{equation}

\section{Connection}
\label{sec:connection}
\begin{figure*}[t!!!]
	\vspace{-1.5ex}
\centerline{
\includegraphics[width = .85\textwidth]{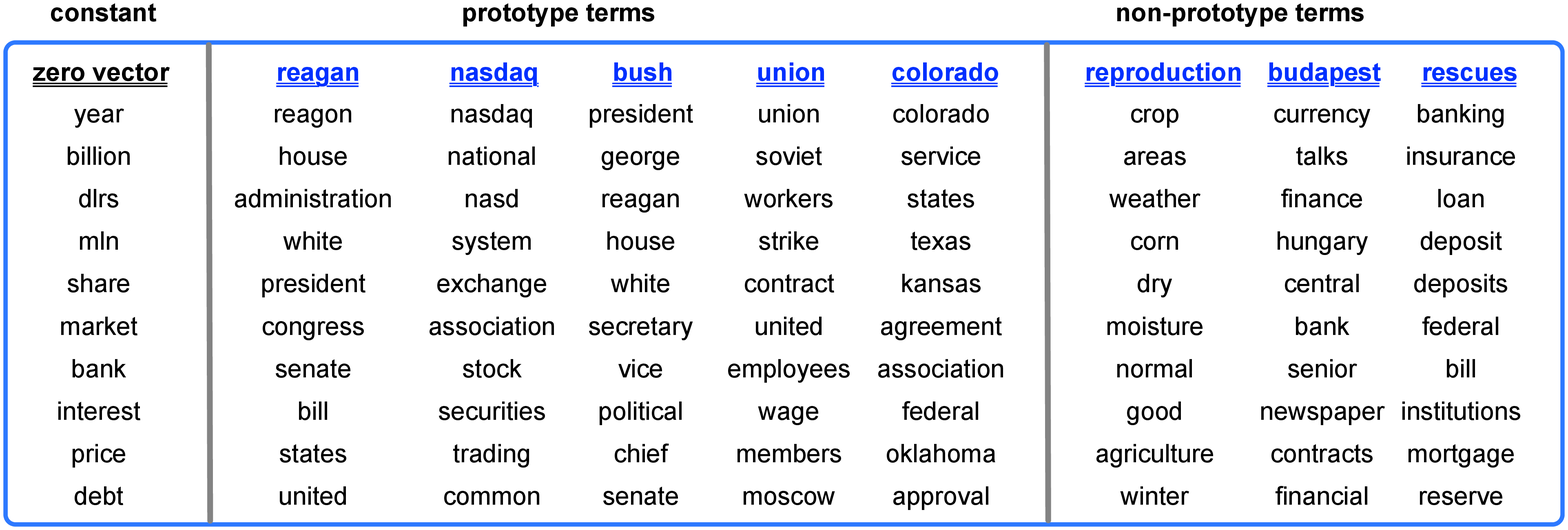}
}
\vspace{-1.5ex}
\caption{Term reconstruction from the Reuters dataset. Each column shows a different input term (\emph{e.g.} ``reagan'', ``nasdaq''), along with the prototype terms reconstructed from this particular input in decreasing order of feature values (top to bottom). The very left column shows the prototype terms generated by an all-empty input document. 
}
\label{fig:reconstruction}
\end{figure*}

\sname{} shares some common elements with previously proposed feature learning algorithms. In this section, we discuss their similarities and differences. 

\textbf{Stacked Denoising Autoencoder (SDA).} In the field of image recognition, the Autoencoder~\cite{hinton1994autoencoders} and the Stacked Denoising Autoencoder (SDA)~\cite{Vincent08extractingand} are widely used to learn better feature representation from raw pixels input. \sname{} shares several core similarities with SDA, which in fact inspired its original development. Similar to \sname{}, SDA first corrupts the raw input, and learns to re-construct it. SDA also stacks several layers together by feeding the output of previous layers as input into sub-sequent layers. 
However, the two algorithms also have substantial differences. The mapping in \sname{} is a linear mapping from input to output (with a sub-sequent application of $\tanh()$), which is solved in closed form. In contrast,  SDA employs non-linear mapping from the input to a hidden layer and then to the output. Instead of a closed-form solution, it requires extensive gradient-descent-type hill-climbing. Further, SDA actually corrupts the input and is trained with multiple epochs over the dataset, whereas \sname{} marginalizes out the corruption. In terms of running time, \sname{} is orders of magnitudes faster than SDA and scales to much higher dimensional inputs~\cite{chen2012msda,xuarxiv2011}.


\textbf{Principle Component Analysis (PCA).} Similar to \sname{}, Principle Component Analysis (PCA)~\cite{jolliffe2002principal} learns a lower dimensional linear space by minimizing the reconstruction error of the original input. For text documents, PCA is widely known through its variant as latent semantic indexing (LSI)~\cite{deerwester1990indexing}. Although both \sname{} and LSI minimize reconstruction errors, the exact optimization is quite different. \sname{} explicitly reconstructs prototype words from corruption, whereas LSI minimizes the reconstruction error after dimensionality reduction. 

\section{Results}
\label{sec:results}

We evaluate our algorithm on \emph{Reuters} and \emph{Dmoz} datasets together with several other algorithms for feature learning. 

\begin{figure}[t!!]
	\vspace{-1.5ex}
\centerline{
\includegraphics[width = .82\columnwidth]{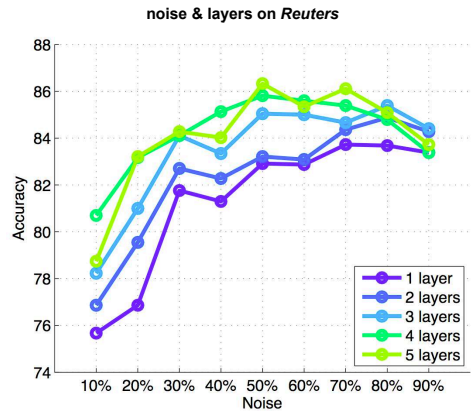}
}
\vspace{-1.5ex}
\caption{Classification accuracy trend on Reuters dataset with different layers and noise levels. 
}
\label{fig:layers}
\end{figure}

\begin{figure*}[t!!!]
	\vspace{-1.5ex}
\centerline{
\includegraphics[width = .80\textwidth]{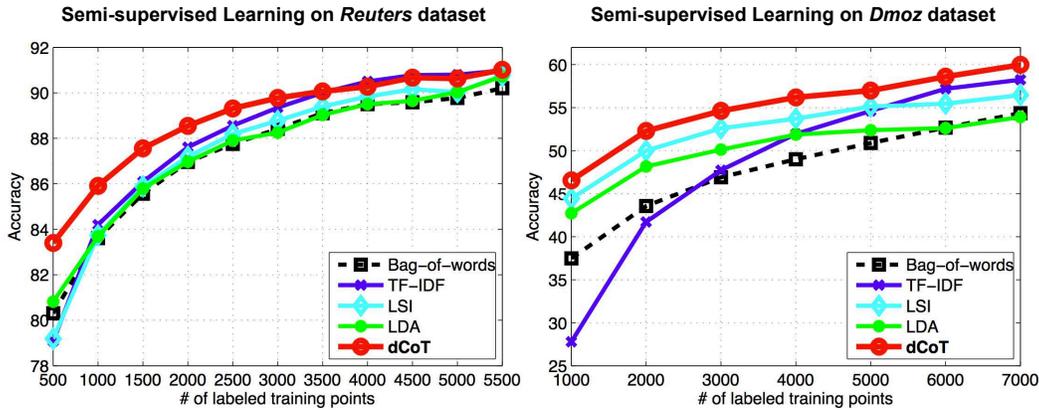}
}
\vspace{-1.5ex}
\caption{Semi-supervised learning results on the Reuters (left) and Dmoz (right) datasets. On both datasets, \sname{} out-performs all other algorithms, especially when the number of labeled inputs is relatively small.  }
\label{fig:semi_supervised}
\end{figure*}

\textbf{Datasets.} The \emph{Reuters-21578} dataset is a collection of documents that appeared on Reuters newswire in $1987$. We follow the convention of~\cite{CWH09}, which removes documents with multiple category labels. The dataset contains 65 categories, and consists of $5946$ training and $2347$ testing documents. Each document is represented by sBoW representation with $18933$ distinct terms. The \emph{Dmoz} dataset is a hierarchical collection of webpage links. The top level of the hierarchy consists of $16$ categories. Following the convention of \cite{NIPS2005_722}, we labeled each input by its top-level category, and remove some low-frequent terms. As a result, the dataset contains $7184$ and $1796$ training and testing points respectively, and each input is represented by the sBoW representation that contains $16498$ distinct terms.  

\textbf{Reconstruction.} Figure~\ref{fig:reconstruction} shows example input terms (essentially one-word documents) and the prototype words that are reconstructed with \sname{} on the Reuters dataset. Each column represents a different input term (\emph{e.g.} a document consisting of only the term ``nasdaq'') and shows the reconstructed prototype terms in decreasing order of their feature values (top to bottom). 
The very left column shows the prototype features generated by an all-empty input document. These features are completely determined by the constant bias, and coincide with the most frequent prototype terms in the whole corpora. For all other columns, we subtract this bias-generated vector to highlight the prototype words generated by the actual word and not the bias. 
As shown in the figure, two trends can be observed. First, prototype terms are reconstructed by other less common and more specific terms. For example, \emph{president} is reconstructed by \emph{reagan} and \emph{bush}, and \emph{stock} is reconstructed by \emph{nasdaq}. Both \emph{reagan} and \emph{bush} are specific terms describing \emph{president}. This trend indicates that \sname{} learns the mapping from rare terms to common terms. Second, context and topics are reconstructed from rarer terms through the recursive re-application. For example, \emph{agriculture} is reconstructed by \emph{reproduction}, indicating that documents containing \emph{reproduction} typically discuss topics related to \emph{agriculture}. This connection indicates that \sname{} also learns the higher order correlations between terms and topics. 

\textbf{Parameter sensitivity.} We also evaluate the effect of different noise level and number of layers (\emph{i.e.} the number of recursive re-applications). Figure~\ref{fig:layers} shows the classification results on Reuters dataset as a function of layers $l$ and noise level $1-p$. After training of \sname{} (on the whole dataset), we randomly select $1,000$ labeled training inputs, train an SVM classifier~\cite{Bishop06} on the new feature representation, and test on the full testing set. 
Two trends emerge: 1. deep layers $l > 1$ improve over a single layered transformation --- supporting our hypothesis that as we recursively re-apply \sname{}, not only the feature representation is enriched, but also the higher order correlations between terms and topics are learned. 2. best results are obtained with a surprisingly high level of noise. 
We explain this trend by the fact that more corruption helps discover more subtle relationships between features and as we operate in the limit, and integrate out all possible corruptions, we can still learn even from substantially shortened documents. 


\textbf{Semi-supervised Experiments.} In many real-world applications, the labeled training inputs are  limited, because labeling usually involves human interaction and  is expensive and time-consuming. However, unlabeled data is usually large and available. In this experiment we evaluate the suitability of \sname{} to take advantage of semi-supervised learning settings. We learn the new feature representation with \sname{} on the full training set (without labels), but train a linear SVM classifier on a small subset of labeled examples. We gradually increase the size of the labeled subset and evaluate on the whole testing set. For any given number of labeled training inputs, we average over five runs (of randomly picked labeled examples). We use the validation set to select the best combination of noise level and the number of layers.

As baselines, we compare against several alternative feature representations, which are all obtained from the full training set, similarly applied to a linear SVM classifier. The most basic baselines are the sBoW representation (with term frequency counts) and TF-IDF~\cite{jones1972statistical}. We compute the TF for each document separately, and obtain the IDF from the whole training set (including labeled and unlabeled data). We then apply the same IDF to the testing set. We also compare against latent semantic indexing (LSI)~\cite{deerwester1990indexing}, for which we further split the training set into training and validation. We use the validation set to find the best parameter (numbers of leading Eigenvectors), and retrain on the whole training set with the best parameter. The new representation is obtained by projecting the sBoW feature space onto the LSI eigenvectors. Finally, we also compare against Latent Dirichlet Allocation (LDA)~\cite{blei2003latent}. Similar to LSI, we use a validation set to find the best parameters, which include the Dirichlet hyper-parameter and the number of topics. The new representation learned from LDA are the topic mixture probabilities. 



The classification results are presented in figure~\ref{fig:semi_supervised}. The graph shows that on both Dmoz and Reuters datasets, \sname{} generally out-performs all other algorithms. This trend is particularly prominent in settings with relatively little labeled training data. 

\begin{table}[t!!]
    \tabcolsep 3.8pt
  	\small
	\vspace{0.25in}
	\center
\begin{tabular}{|l|c|c|c|c|}
	\hline {\bf Datasets} &  {\bf TF-IDF} & {\bf LSI} & {\bf LDA} & {\bf \sname{}} \\
	\hline
	\hline Reuters & 1s & 51m & 3h10m & 2m \\
	\hline Dmoz & 1s & 1h38m & 9h1m & 3m \\  
	\hline
\end{tabular}
\caption{Running time required for unsupervised feature learning with different algorithms.}
~\label{table:timing}
\end{table}

\textbf{Running time.} Table~\ref{table:timing} compares the running times for feature learning with different algorithms. All timings are performed on a desktop with dual Intel$^{TM}$Six Core Xeon X5650 2.66GHz processors. Compared to LDA and LSI, the timing results show a three orders of magnitude speed-up on two datasets, reducing the feature learning time from several hours to a few minutes.

\section{Conclusion}
\label{sec:conclusion}
In this paper we present \sname{}, an algorithm that efficiently learns a better feature representation for sBoW document data. Specifically, \sname{} learns a mapping from high dimensional sparse to low dimensional dense representations by translating rare to common terms. Recursive re-application of \sname{} on its own output results in the discovery of higher order topics from raw terms. On two standard benchmark document classification datasets we demonstrate that our algorithm achieves state-of-the-art results with very high reliability in semi-supervised settings. 

\section*{Acknowledgment}
This material is based upon work supported by the National Science Foundation under Grant No. 1149882. Any opinions, findings, and conclusions or recommendations expressed in this material are those of the author(s) and do not necessarily reflect the views of the National Science Foundation.

\bibliographystyle{IEEEtran}
\bibliography{IEEEabrv,msda_dm}

\end{document}